\documentclass[oupdraft]{bio}
\usepackage{url}
\usepackage{booktabs}

\begin{document}

\title{Estimating intracluster correlation for ordinal data }

\author{Benjamin W. Langworthy$^{1,2,\ast}$, Zhaoxun Hou$^1$, Gary C. Curhan$^{2,3,4,5}$, Sharon G. Curhan$^{3,4}$,\\
Molin Wang$^{1,2,3}$\\[4pt]
\textit{$^1$ Department of Biostatistics, Harvard TH Chan School of Public Health, Boston, MA, USA\\
$^2$ Department of Epidemiology, Harvard TH Chan School of Public Health, Boston, MA, USA\\
$^3$ Department of Medicine, Channing Division of Network Medicine, Brigham and Women's Hospital, Boston, MA, USA\\
$^4$ Harvard Medical School, Boston, MA, USA\\
$^5$ Renal Division, Brigham and Women's Hospital, Boston, MA, USA}\\[2pt]
{blangworthy@hsph.harvard.edu}}







{B. W. Langworthy and others}
{Intraculster correlation for ordinal data}

\maketitle


\begin{abstract}{\textbf{Purpose:} In this paper we consider the estimation of intracluster correlation for ordinal data. We focus on pure-tone audiometry hearing threshold data, where thresholds are measured in 5 decibel increments. We estimate the intracluster correlation for tests from iPhone-based hearing assessment application as a measure of test/retest reliability.\\
\textbf{Methods:} We present a method to estimate the intracluster correlation using mixed effects cumulative logistic and probit models, which assume the outcome data are ordinal. This contrasts with using a mixed effects linear model which assumes that the outcome data are continuous.\\
\textbf{Results:} In simulation studies we show that using a mixed effects linear model to estimate the intracluster correlation for ordinal data results in a negative finite sample bias, while using mixed effects cumulative logistic or probit models reduces this bias. The estimated intracluster correlation for the iPhone-based hearing assessment application is higher when using the mixed effects cumulative logistic and probit models compared to using a mixed effects linear model.\\
\textbf{Conclusion:} When data are ordinal, using mixed effects cumulative logistic or probit models reduces the bias of intracluster correlation estimates relative to using a mixed effects linear model.
}{Test/retest reliability, Reliability and validity, Pure-tone audiometry, Intracluster correlation} 
\end{abstract}

\maketitle
\section*{List of abbreviations and acronyms}
\textbf{dB:} Decibel\\
\textbf{Hz:} Hertz\\
\textbf{ICC:} Intracluster correlation\\
\textbf{NHS II:} Nurses' Health Study II

\section{Introduction}
Intracluster correlation (ICC) is frequently used to measure test/retest reliability, and can be used as one tool to measure the quality of newly developed testing procedures. We consider audiometric hearing threshold data from an iPhone-based hearing assessment application. Pure-tone audiometry assesses the quietest tone, “hearing threshold,” that can be detected at specific frequencies in each ear. If tests are conducted on the same subject in a short enough time frame we expect the true underlying hearing threshold to remain constant for each test. Therefore a reliable test should have a high ICC, indicating strong correlation among the hearing thresholds of the same ear. 
\par ICC can be estimated using a linear mixed effects model which assumes the random effects and random error terms are normally distributed. However, pure tone audiometry tests typically measure hearing thresholds in 5 decibel (dB) increments. Therefore, hearing threshold data are ordinal rather than continuous. In simulations we show that for finite samples, treating the data as continuous, rather than ordinal, biases the estimates of the ICC toward zero. A framework for estimating the ICC for generalized linear mixed effects models has been proposed in \citet{nakagawa2017coefficient} and \citet{nakagawa2013general,nakagawa2010repeatability}. We extend this framework to cumulative link models for ordinal data. We show that using this ordinal model framework can obtain more accurate ICC estimates than naively treating ordinal data as continuous.  
\section{Modelling framework}
 Let $Y^*_{ij}$ denote the true hearing threshold for the $jth$ measurement for the $ith$ ear on a continuous scale. Assume that $Y^*_{ij}$ follows a linear mixed effects model,
\begin{equation}
    Y^*_{ij} = X_{ij}^T\beta^* + b^*_i + \epsilon^*_{ij},\label{eq:singclust}
\end{equation}
where $X_{ij}$ is a vector of covariates, $\beta^*$ is a vector of fixed effect parameters, $b^*_i \sim N(0,\sigma^2_{b^*})$ is an ear specific random effect, and $\epsilon^*_{ij} \sim N(0,\sigma^2_{\epsilon^*})$ is a normally distributed random error term. We focus on the adjusted ICC, which is
\begin{equation}
    ICC^{adj}_{Y^*_{ij}} = \frac{\sigma_{b*}^2}{\sigma_{b*}^2 + \sigma_{\epsilon *}^2}\text{(\citealp{nakagawa2017coefficient})}.\nonumber 
    \end{equation}
For hearing threshold data, instead of observing $Y^*_{ij}$ we observe $Y_{ij}$, which has $K$ categories. In order to categorize $Y_{ij}$ we assume there are $K+1$ different cutpoints, $\xi_k$, where $Y_{ij} = k$ if $\xi_{k-1} < Y^*_{ij} < \xi_{k}$. In this case $\xi_0 = -\infty$ and $\xi_K = \infty$, and all the other cutpoints are finite values put into ascending order. This can be written as a mixed effects cumulative probit model,
\begin{equation}
     P(Y_{ij} \le k) = P(Y^*_{ij} \le \xi_k) = \Phi\left(\frac{\xi_k - X_{ij}^T\beta^* - b_i^*}{\sigma_{\epsilon *}}\right) = \Phi\left(\frac{\xi_k}{\sigma_{\epsilon *}} -  \frac{X_{ij}^T\beta^*}{\sigma_{\epsilon *}} - \frac{b_i^*}{\sigma_{\epsilon *}}\right),\nonumber
\end{equation}
 where the fixed effects coefficients are $\beta = \beta^*/\sigma_{\epsilon *}$, and the random effects are $b_i = b^*_i/\sigma_{\epsilon *}$, with $b_i \sim N(0,\sigma_b^2)$, and $\sigma_b^2 = \sigma_{b*}^2/\sigma_{\epsilon *}^2$ (\citealp{agresti2003categorical,hedeker1994random}). We can recover the adjusted ICC for the latent variable, $Y^*_{ij}$, as
\begin{equation}
    ICC_{Y_{ij}^*prob}^{adj} = \frac{\sigma_b^2}{\sigma_b^2 + 1 }.\nonumber
    \end{equation}
If we assume that $\epsilon^*_{ij}$ has a logistic distribution, the same method would imply a mixed effects cumulative logistic model. In this case, when calculating the adjusted ICC, we replace the one in the denominator with $\pi^2/3$, the variance of a standard logistic distribution, 
\begin{equation}
    ICC_{Y_{ij}^*log}^{adj} = \frac{\sigma_b^2}{\sigma_b^2 + \pi^2/3 }.\nonumber
    \end{equation}
\par Above we treat each ear as it's own sampling unit. It is also possible to think of each subject as a sampling unit with a random effect for subject and individual ears. Assume that $Y^*_{ikj}$ is the underlying true hearing threshold on a continuous scale of the $jth$ measurement for the $ith$ subject in the $kth$ ear. Assume $Y^*_{ikj}$ follows the linear mixed effects model,
\begin{equation}
    Y^*_{ikj} = X_{ikj}^T \beta^* + b_i^* + c_{ik}^* + \epsilon_{ikj}^*, \label{eq:multiclust}
\end{equation}
where $b_i^*\sim N(0,\sigma^2_{b^*})$ is a subject specific random effect, $c_{ik}^* \sim N(0,\sigma^2_{c^*})$ is an ear specific random effect, and $\epsilon_{ikj}^*\sim N(0,\sigma^2_{\epsilon^*})$ is a random error term. The adjusted ICC is
\begin{equation}
    ICC^{adj}_{Y^*_{ikj}} = \frac{\sigma_{b*}^2+\sigma_{c*}^2}{\sigma_{b*}^2 +\sigma_{c*}^2+ \sigma_{\epsilon *}^2}.\nonumber
    \end{equation}
    Using the same methods as the single random effect model, the adjusted ICC based on the mixed effects cumulative probit and logistic models can be written as
    \begin{equation}
        ICC_{Y_{ikj}^*prob/log}^{adj} = \frac{\sigma_b^2 + \sigma_c^2}{\sigma_b^2 + \sigma_c^2 + m},\nonumber
    \end{equation}
    where $m=1$ for the mixed effects cumulative probit model, $m=\pi^2/3$ for the mixed effects cumulative logistic model, $\sigma_b = \sigma_{b^*}^2/\sigma_{\epsilon^*}^2$, and $\sigma_c = \sigma_{c^*}^2/\sigma_{\epsilon^*}^2$.

\section{Simulation study}
We use simulation studies to compare the finite sample performance of the adjusted ICC estimates using mixed effects cumulative probit or logistic models to those using mixed effects linear regression, when the simulated data are ordinal. First we consider a single level of clustering, where the continuous outcome $Y^*_{ij}$ is simulated according to the mixed effects linear model from Equation (\ref{eq:singclust}),
where $X_{ij}$ is a scalar covariate, $\beta^*$ is the fixed effect parameter, $b^*_i$ is an ear specific random effect, and $\epsilon^*_{ij}$ is a random error term. We simulate 35 groups with 5 observations per group, such that $i=1,\ldots,35$ and $j=1,\ldots,5$. We simulate  $X \sim N(0,1)$, $\beta^* = 1$, and $b^*_i \sim N(0,4)$. For the random error term we simulate both $\epsilon^* \sim N(0,1)$ and $\epsilon^* \sim \text{Logistic}(0,2\sqrt{3}/\pi)$, with the former corresponding to a properly specified cumulative probit model and the latter corresponding to a properly specified cumulative logistic model after using cutpoints to define an ordinal variable based on $Y^*_{ij}$. In both cases the true adjusted ICC is 0.8. In order to define the ordinal variable, $Y_{ij}$, we define the cutpoints, $\xi_k$, to be all even integers including zero. This results in 7-9 categories for most simulations. In order to estimate the ICC we fit mixed effects cumulative probit and logistic models model and use the methods described in Section 2. We also use the naive method, in which we fit a mixed effects linear model with the ordinal variable, $Y_{ij}$, as the outcome. 
\par In addition to the single level of clustering, we also consider simulations using two levels of clustering where we simulate $Y^*_{ikj}$ according to the linear mixed effects model from Equation (\ref{eq:multiclust}),
where $b_i^*$ is a subject specific random effect, $c_{ik}^*$ is an ear specific random effect, and $\epsilon_{ikj}^*$ is a random error term. We simulate  such that $i=1,\ldots,35$, $k=1,2$ and $j=1,\ldots,5$. This is equivalent to a study where there are 35 subjects with 2 ears each, and 5 measurements per ear. We define $X$, $\beta^*$ and $\epsilon^*$ in the same way as the single level clustering example. We simulate $b^*_{i}\sim N(0,2)$, and $c^*_{ik}\sim N(0,2)$, so that the true ICC is again 0.8. Again, we define the ordinal variable $Y_{ikj}$ by setting the cutpoints, $\xi_k$, to be all even integers including zero. As with the single clustering simulations we estimated the ICC using methods described in Section 2. We compare using mixed effects cumulative logistic and probit models to a mixed effects linear model, with $Y_{ikj}$ as the outcome. 
\par For the single random effect setup we note that $ICC_{Y_{ij}^*prob}^{adj}$ and $ICC_{Y_{ij}^*log}^{adj}$ are both transformations of $\sigma_b^2$. Therefore we can calculate the confidence interval for $ICC_{Y_{ij}^*prob}^{adj}$ and $ICC_{Y_{ij}^*log}^{adj}$ by transforming the confidence interval for $\sigma_b^2$, which can be obtained using the likelihood root statistic from the profile likelihood \cite{christensen2015analysis,christensen2018cumulative}. In the multiple random effect setting we can obtain a confidence interval for $ICC_{Y_{ijk}^*prob}^{adj}$ and $ICC_{Y_{ijk}^*log}^{adj}$ by applying the delta method based on the estimated variance/covariance matrix of $\sigma_b^2$ and $\sigma_c^2$. We include an example of how to estimate the adjusted ICC and confidence intervals for all methods at https://github.com/blangworthy/ICCordinal. 
 \par The empirical bias, standard deviation, and $95\%$ confidence interval coverage rate based on 1000 simulations for both the single and multi-level clustering setups are reported in Table \ref{tab:simbiassdcr}. The adjusted ICC estimates using the mixed effect cumulative logistic and probit models both have bias close to zero while the naive estimator unerestimates the adjusted ICC. In a small number of simulated data sets (1.5\% or less) the confidence intervals could not be estimated because the cumulative logistic and probit models are unstable when the estimated ICC is one or very close to one. When the ICC is one the variance of $\epsilon_{ij}^*$ or $\epsilon_{ijk}^*$ is zero, and the cumulative logistic or probit models will not be well defined due to division by zero. We exclude the simulated data sets where the confidence intervals could not be estimated when calculating the coverage rate. The confidence intervals have close to the desired coverage rate when using the cumulative logistic or probit models.

 \begin{table}[htbp]
   \caption{Bias (SD) and $95\%$ confidence interval coverage rate of adjusted ICC estimates for single and multi-level clustering using three different methods for estimation}
    \begin{tabular}{lrrrrrrrr}
     \toprule
           & \multicolumn{4}{c}{Single level clustering$^3$} & \multicolumn{4}{c}{Multilevel Clustering$^3$} \\
           & \multicolumn{2}{c}{Normal Error$^4$} & \multicolumn{2}{c}{Logistic  Error$^4$} & \multicolumn{2}{c}{Normal Error} & \multicolumn{2}{c}{Logistic Error} \\
           & Bias (SD)$^1$ & CR$^2$    & Bias (SD) & CR    & Bias (SD) & CR    & Bias (SD) & CR \\
     \midrule
     Adj ICC Probit & -0.01 (0.05) & 0.95  & -0.01 (0.06) & 0.92  & -0.01 (0.04) & 0.93  & -0.01 (0.04) & 0.92 \\
     Adj ICC Logistic & -0.01 (0.06) & 0.95  & -0.01 (0.06) & 0.93  & -0.01 (0.04) & 0.94  & -0.01 (0.04) & 0.92 \\
     Adj ICC Naive & -0.06 (0.06) & 0.89  & -0.06 (0.06) & 0.87  & -0.06 (0.04) & 0.80  & -0.06 (0.04) & 0.81 \\
     \bottomrule
     
     \multicolumn{9}{l}{True ICC = 0.8}\\
     \multicolumn{9}{l}{$^1$Bias is the estimated ICC minus the true ICC; SD is the empirical standard deviation of ICC }\\
     \multicolumn{9}{l}{estimates  over all simulation replicates.}\\
     \multicolumn{9}{l}{$^2$CR is coverage rate and is the proportion of 95\% confidence intervals that contain the true ICC; }\\
     \multicolumn{9}{l}{confidence intervals for adjusted ICC using probit and logistic models in single level clustering use}\\ 
     \multicolumn{9}{l}{profile confidence interval.}\\
     \multicolumn{9}{l}{$^3$Single level clustering setup includes 35 subjects and 5 measures per subject. Multilevel clustering }\\
     \multicolumn{9}{l}{setup includes 35 subjects, 2 ears per subject, and 5 measures per ear.}\\
     \multicolumn{9}{l}{$^4$When error has normal distribution probit model is properly specified. When error has logistic }\\
     \multicolumn{9}{l}{distribution logistic model is properly specified.}
     \end{tabular}%
   \label{tab:simbiassdcr}%
 \end{table}%

\section{Illustrative example}\label{sec:illus}
To illustrate our method, we analyze data from 31 women from the Nurses’ Health Study II (NHS II) (\citealp{bao2016origin}), who self-administered home hearing tests using the Decibel Therapeutics iPhone hearing assessment application. The home hearing test assessed pure-tone air conduction hearing thresholds (in 5 dB steps) across the conventional range of frequencies between 500 and 8000 hertz (Hz) in each ear. In the study 21 (68\%) of the participants had measurements from four repeated home hearing tests; 27 (87\%) had measurements from at least two repeated tests. The average number of home hearing assessments was 3.38. The results for single-ear and both-ear adjusted ICCs for hearing threshold measurements obtained at 8000 Hz are reported in Table \ref{tab:realexample}, with results for other frequencies reported in the supplementary materials. The adjusted ICC estimated using the mixed effects cumulative logistic model is the highest, and the adjusted ICC estimated using the mixed effects linear model is the lowest. This is consistent with simulation results, in which the naive estimator using the mixed effects linear model has a negative finite sample bias. The adjusted ICC is above 0.85 in all cases, indicating strong test/retest reliability.
\begin{table}[htbp]

  \caption{Single-ear and both-ear adjusted ICCs and 95\% Wald and profile confidence intervals for 8000 Hz}
    \begin{tabular}{lrrrrrr}
    \toprule
          & \multicolumn{2}{c}{Cumulative Probit} & \multicolumn{2}{c}{Cumulative Logistic} & \multicolumn{2}{c}{Naive } \\
          & Adj ICC   & 95\% CI$^1$ & Adj ICC   & 95\% CI & Adj ICC   & 95\% CI \\
          \midrule
    Frequency/Ear & &&&&& \\
    \midrule
    8000 Hz/Right & 0.92  & (0.84,0.96) & 0.94  & (0.87,0.98) & 0.89  & (0.81,0.96) \\
    8000 Hz/Left & 0.86  & (0.75,0.93) & 0.91  & (0.82,0.96) & 0.86  & (0.78,0.95)  \\
    8000  Hz/Both & 0.89  & (0.83,0.95) & 0.93  & (0.88,0.97) & 0.88  &  (0.82,0.94) \\
    \bottomrule
    
    \multicolumn{7}{l}{$^1$Confidence intervals for single-ear adjusted cumulative logistic and probit models are }\\
    \multicolumn{7}{l}{estimated using transformed profile confidence intervals, all other confidence intervals}\\
    \multicolumn{7}{l}{estimated using the delta method}\\
    \multicolumn{7}{l}{Mixed effects model included mean ambient noise as covariate}
    \end{tabular}%
  \label{tab:realexample}%
\end{table}%

\section{Discussion}
We show that when data are ordinal, using a linear mixed effects model which treats data as continuous can underestimate the adjusted ICC. A more precise estimate of the adjusted ICC can be obtained using a mixed effects cumulative logistic or probit model. We apply our methods to hearing threshold data obtained from participants in the NHS II. In this example the estimated adjusted ICC using mixed effects cumulative logistic or probit models is higher than the estimated adjusted ICC using a linear mixed effects model. Example code for estimating the adjusted ICC can be found at https://github.com/blangworthy/ICCordinal. 

\section{Supplementary Material}
 The supplementary materials include a description of the data set used in the illustrative example and results for additional frequencies.
 
 \section{Funding}
 This work was supported by the National Institute Health grants R01 DC017717 and U01 DC010811.
\bibliographystyle{biorefs}
\bibliography{iccordinal}

\begin{thebibliography}{99}

\bibitem[Agresti(2003)Agresti]{agresti2003categorical}
\textsc{Agresti, Alan}. (2003).
\newblock {\em Categorical data analysis\/}, Volume 482. John Wiley \& Sons.

\bibitem[Bao \emph{and others}(2016)Bao, Bertoia, Lenart, Stampfer, Willett,
  Speizer and Chavarro]{bao2016origin}
\textsc{Bao, Ying, Bertoia, Monica~L, Lenart, Elizabeth~B, Stampfer, Meir~J,
  Willett, Walter~C, Speizer, Frank~E and Chavarro, Jorge~E}. (2016).
\newblock Origin, methods, and evolution of the three nurses’ health studies.
\newblock {\em American journal of public health\/}~\textbf{106}(9),
  1573--1581.

\bibitem[Christensen(2015)Christensen]{christensen2015analysis}
\textsc{Christensen, Rune Haubo~B}. (2015).
\newblock Analysis of ordinal data with cumulative link models—estimation
  with the r-package ordinal.
\newblock {\em R-package version\/}~\textbf{28}, 406.

\bibitem[Christensen(2018)Christensen]{christensen2018cumulative}
\textsc{Christensen, Rune Haubo~B}. (2018).
\newblock Cumulative link models for ordinal regression with the r package
  ordinal.
\newblock {\em Submitted in J. Stat. Software\/}.

\bibitem[Hedeker and Gibbons(1994)Hedeker and Gibbons]{hedeker1994random}
\textsc{Hedeker, Donald and Gibbons, Robert~D}. (1994).
\newblock A random-effects ordinal regression model for multilevel analysis.
\newblock {\em Biometrics\/}, 933--944.

\bibitem[Nakagawa \emph{and others}(2017)Nakagawa, Johnson and
  Schielzeth]{nakagawa2017coefficient}
\textsc{Nakagawa, Shinichi, Johnson, Paul~CD and Schielzeth, Holger}. (2017).
\newblock The coefficient of determination $r^2$ and intra-class correlation
  coefficient from generalized linear mixed-effects models revisited and
  expanded.
\newblock {\em Journal of the Royal Society Interface\/}~\textbf{14}(134),
  20170213.

\bibitem[Nakagawa and Schielzeth(2010)Nakagawa and
  Schielzeth]{nakagawa2010repeatability}
\textsc{Nakagawa, Shinichi and Schielzeth, Holger}. (2010).
\newblock Repeatability for gaussian and non-gaussian data: a practical guide
  for biologists.
\newblock {\em Biological Reviews\/}~\textbf{85}(4), 935--956.

\bibitem[Nakagawa and Schielzeth(2013)Nakagawa and
  Schielzeth]{nakagawa2013general}
\textsc{Nakagawa, Shinichi and Schielzeth, Holger}. (2013).
\newblock A general and simple method for obtaining $r^2$ from generalized
  linear mixed-effects models.
\newblock {\em Methods in ecology and evolution\/}~\textbf{4}(2), 133--142.

\end{thebibliography}


\begin{thebibliography}{99}

\bibitem[Bao \emph{and others}(2016)Bao, Bertoia, Lenart, Stampfer, Willett,
  Speizer and Chavarro]{bao2016origin}
\textsc{Bao, Ying, Bertoia, Monica~L, Lenart, Elizabeth~B, Stampfer, Meir~J,
  Willett, Walter~C, Speizer, Frank~E and Chavarro, Jorge~E}. (2016).
\newblock Origin, methods, and evolution of the three nurses’ health studies.
\newblock {\em American journal of public health\/}~\textbf{106}(9),
  1573--1581.

\end{thebibliography}

\end{document}


\title{Estimating intracluster correlation for ordinal data supplementary materials}

\author{Benjamin W. Langworthy$^{1,2,\ast}$, Zhaoxun Hou$^1$, Gary C. Curhan$^{2,3,4,5}$, Sharon G. Curhan$^{3,4}$,\\
Molin Wang$^{1,2,3}$\\[4pt]
\textit{$^1$ Department of Biostatistics, Harvard TH Chan School of Public Health, Boston, MA, USA\\
$^2$ Department of Epidemiology, Harvard TH Chan School of Public Health, Boston, MA, USA
\\
$^3$ Department of Medicine, Channing Division of Network Medicine, Brigham and Women's Hospital, Boston, MA, USA\\
$^4$ Harvard Medical School, Boston, MA, USA\\
$^5$ Renal Division, Brigham and Women's Hospital, Boston, MA, USA}\\[2pt]
{blangworthy@hsph.harvard.edu}}







\markboth%
{B. W. Langworthy and others}
{Intraculster correlation for ordinal data}

\maketitle

\renewcommand{\tablename}{S}
\section{Illustrative example}
In Section 4 of the main text we analyze data from 31 women from the Nurses’ Health Study II (NHS II) (\citealp{bao2016origin}), who self-administered home hearing tests using the Decibel Therapeutics iPhone hearing assessment application. The results for single-ear and both-ear adjusted ICCs for hearing threshold measurements obtained at 8000 hertz (Hz) are presented in Section 4 of the main text. Table S\ref{tab:realexamplesupp} presents the same results obtained at 500, 1000, 2000, 3000, 4000 and 6000 Hz. Across all frequencies the adjusted ICC estimated using a mixed effects cumulative logistic model tends to be highest, followed by the adjusted ICC estimated using a mixed effects cumulative probit model, and the naive method using a mixed effects linear model. One thing we note is that for the adjusted ICC using both ears at 4000 Hz cannot be estimated using the cumulative logistic model. This is due to the issue with the estimated adjusted ICC being one causing issues due to division by zero, as discussed in Sections 3 of the main text. There are a number of frequencies where the confidence intervals cannot be estimated due to a similar issue.

\begin{table}[ht]
\centering
\caption{Single-ear adjusted ICCs and 95\% Wald and profile confidence intervals for all 500, 1000, 2000, 3000, 4000, and 6000 Hz}
\begin{tabular}{lrrrrrr}
  \hline
    & \multicolumn{2}{c}{Cumulative Probit} & \multicolumn{2}{c}{Cumulative Logistic} & \multicolumn{2}{c}{Naive } \\
          & Adj ICC   & 95\% CI$^1$ & Adj ICC   & 95\% CI & Adj ICC   & 95\% CI \\
          \midrule
    Frequency/Ear & &&&&& \\
  \hline
500 Right & 0.80 & (0.63,0.90) & 0.81  & NA & 0.80 & (0.69,0.92) \\ 
  500/Left & 0.92  & NA & 0.93  & (0.84,0.97) & 0.89 & (0.83,0.96) \\ 
  1000/Right & 0.83 & (0.69,0.91) & 0.86  & (0.73,0.93) & 0.77 & (0.64,0.90) \\ 
  1000/Left & 0.77  & (0.57,0.88) & 0.79  & NA & 0.73 & (0.57,0.88) \\ 
  2000/Right & 0.80 & (0.65,0.90) & 0.90  & NA & 0.78 & (0.66,0.91) \\ 
  2000/Left & 0.60  & (0.34,0.79) & 0.75 & (0.52,0.88) & 0.57 & (0.36,0.79) \\ 
  3000/Right & 0.78  & (0.60,0.89) & 0.83  & (0.67,0.92) & 0.75 & (0.61,0.89) \\ 
  3000/Left & 0.99  & NA & 0.99  & NA & 0.97 & (0.95,0.99) \\ 
  4000/Right & 0.75  & (0.56,0.88) & 0.85  & (0.70,0.94) & 0.84 & (0.74,0.94) \\ 
  4000/Left & 0.99  & NA & 1.00  & NA & 0.98 & (0.96,0.99) \\ 
  6000/Right & 0.85  & (0.72,0.93) & 0.89  & (0.78,0.95) & 0.80 & (0.69,0.92) \\ 
  6000/Left & 0.92  & NA & 0.93  & (0.83,0.97) & 0.93 & (0.88,0.98) \\ 
   500/Both & 0.85 & (0.78,0.93) & 0.87 & (0.79,0.94) & 0.85 & (0.77,0.92) \\ 
  1000/Both & 0.81 & (0.71,0.91) & 0.84 & (0.75,0.93) & 0.76 & (0.65,0.87) \\ 
  2000/Both & 0.73 & (0.60,0.85) & 0.84 & (0.76,0.92) & 0.70 & (0.57,0.83) \\ 
  3000/Both & 0.87 & (0.80,0.93) & 0.92 & (0.88,0.97) & 0.85 & (0.78,0.92) \\ 
  4000/Both & 0.92 & (0.87,0.96) & NA & NA & 0.91 & (0.87,0.96) \\ 
  6000/Both & 0.88 & (0.81,0.95) & 0.90 & (0.84,0.96) & 0.85 & (0.78,0.93) \\ 
  \hline
  \multicolumn{7}{l}{$^1$Confidence intervals for single-ear adjusted cumulative logistic and probit models are }\\
    \multicolumn{7}{l}{estimated using transformed profile confidence intervals, all other confidence intervals}\\
    \multicolumn{7}{l}{estimated using the delta method}\\
    \multicolumn{7}{l}{Mixed effects model included mean ambient noise as covariate}
\end{tabular}
\label{tab:realexamplesupp}%
\end{table}

\bibliographystyle{biorefs}
\bibliography{iccordinal}